\documentclass[fleqn,usenatbib]{mnras}
\usepackage{newtxtext,newtxmath}
\usepackage[T1]{fontenc}
\usepackage{ae,aecompl}
\usepackage{natbib}
\usepackage{graphicx}	% Including figure files
\usepackage{amsmath}	% Advanced maths commands
\usepackage{amssymb}	% Extra maths symbols

\title[Scattering of Strahl Electrons in the Solar Wind]{Scattering of Strahl Electrons in the Solar Wind between 0.3 and 1 au: Helios Observations}

\author[L. Ber\v{c}i\v{c} et al.]{L. Ber\v{c}i\v{c}$^{1, 2}$\thanks{E-mail: laura.bercic@obspm.fr}, 
M. Maksimovi\'{c} $^{1}$, S. Landi $^{2,3}$, L. Matteini $^{1}$ \\
$^{1}$LESIA, Observatoire de Paris, PSL Research University, CNRS, UPMC Université Paris 6, Université Paris-Diderot, 5 Place \\Jules Janssen, F-92190 Meudon, France\\
$^{2}$Physics and Astronomy Department, University of Florence, Via Giovanni Sansone 1, I-50019 Sesto Fiorentino, Italy\\
$^{3}$INAF - Osservatorio Astrofisico di Arcetri, Largo Enrico Fermi 5, I-50125 Firenze, Italy
}

\date{Accepted 2019 April 4. Received 2019 April 1; in original form 2018 November 17}

\pubyear{2015}

\begin{document}
\label{firstpage}
\pagerange{\pageref{firstpage}--\pageref{lastpage}}
\maketitle

% Abstract of the paper
\begin{abstract}
Electron velocity distribution functions in the solar wind according to standard models consist of 4 components, of which 3 are symmetric - the core, the halo, and the superhalo, and one is magnetic field-aligned, beam-like population, referred to as the strahl. We analysed in-situ measurements provided by the two Helios spacecrafts to study the behaviour of the last, the strahl electron population, in the inner Solar system between 0.3 and 1 au. The strahl is characterised with a pitch-angle width (PAW) depending on electron energy and evolving with radial distance. We find different behaviour of the strahl electrons for solar wind separated into types by the core electron beta parallel value ($\beta_{ec\parallel}$). For the low-$\beta_{ec\parallel}$ solar wind the strahl component is more pronounced, and the variation of PAW is electron energy dependent. At low energies a slight focusing over distance is observed, and the strahl PAW measured at 0.34 au agrees with the width predicted by a collisionless focusing model. The broadening observed for higher-energy strahl electrons during expansion can be described by an exponential relation, which points toward an energy dependent scattering mechanism. In the high-$\beta_{ec\parallel}$ solar wind the strahl appears broader in consistence with the high-$\beta_{ec\parallel}$ plasma being more unstable with respect to kinetic instabilities. Finally we extrapolate our observations to the distance of 0.16 au, predicting the strahl PAWs in the low-$\beta_{ec\parallel}$ solar wind to be $\sim$ 29$^o$ for all energies, and in the high-$\beta_{ec\parallel}$ solar wind a bit broader, ranging between 37$^o$ and 65$^o$.
\end{abstract}

% Select between one and six entries from the list of approved keywords.
% Don't make up new ones.

%%%%%%%%%%%%%%%%%%%%%%%%%%%%%%%%%%%%%%%%%%%%%%%%%%

%%%%%%%%%%%%%%%%% BODY OF PAPER %%%%%%%%%%%%%%%%%%

\section{Introduction}

Electrons as the lighter constituents of the solar wind are the carriers of the heat flux and therefore play an important role in the energy balance during the solar wind expansion. Electron velocity distribution functions (VDF) are highly non-thermal and can be divided into 4 components: a core, a thermal and dense population well represented by a Maxwellian function, a halo with a higher temperature and exhibiting strong high-energy tails, an even hotter superhalo spanning from a few to a few hundred keV, and a magnetic field aligned component, called a strahl \citep{Feldman1975,Hammond1996,Maksimovic1997b, Lin1998, Maksimovic2005, Wang2012,Graham2017}.

Strahl electrons can propagate in a positive or negative magnetic field direction, but generally away from the Sun \citep{Feldman1978,Pilipp1987a}. Bi-directional strahls have also been observed and serve as indicators of certain magnetic field structures, like magnetic field loops and magnetic clouds \citep{Gosling1987}. 

It is commonly believed that these anti-sunward field-aligned electrons originate from the hot solar corona, escaping from a thermal VDF and focusing around the magnetic field as they conserve their magnetic moments \citep{Feldman1975,Pierrard1999, Salem2007}. The formation of the strahl from a thermal population during the spherical expansion was simulated by \citet{Landi2012} using a fully kinetic model including Coulomb collisions. However, it was shown with particle-in-cell simulations that strahl could also be created by a resonant interaction of halo electrons with whistler-mode waves generated by electron core anisotropy \citep{Seough2015}. The question of the origin of strahl electrons as well as other non-thermal components of electron VDF awaits for new theoretical and observational studies, soon fortified by the two upcoming solar missions: Parker Solar Probe and Solar Orbiter.

The properties of strahl population, and its evolution during the expansion have been shown on the basis of various near-Earth, and interplanetary in-situ observations. The theoretically predicted focusing effect during the radial expansion was not observed. On the opposite, widening of strahl VDF with distance from the Sun has been reported by \citet{Hammond1996} using Ulysses data (1 - 3.5 au), and \citet{Graham2017} using Cassini data (1 - 6 au). The authors of the later state that the strahl ceases to exist at distances larger than 5.5 au as it is most likely completely scattered into the halo population. This hypothesis agrees with the study of \citet{Maksimovic2005} \& \citet{Stverak2009}, showing a decrease in relative density of strahl component with radial distance, but an increase of the halo density. \citet{Stverak2009} find the same tendency in both the slow and the fast solar wind between 0.3 and 4 au using data from Helios, Cluster and Ulysses missions.

The strahl is more pronounced and narrower in the fast wind as oppose to the slow wind, where it appears less dense, broader, and sometimes even not present at 1 au \citep{Fitzenreiter1998} \& \citep{Gurgiolo2017}. 

Studying the variation of the strahl pitch-angle width with electron energy might reveal which scattering mechanisms are at work at different radial distances, and for different solar wind types. Both increasing, and decreasing trends were observed so far. \citet{Kajdic2016} find anti-correlation between pitch-angle width and electron energy, which gets broken for a certain energy range at times of observed whistler-mode wave activity. Their analysis includes mostly the slow solar wind at 1 au (Cluster observations). Particle-in-cell simulations provided by \citet{Saito2007} confirm that strahl scattered by whistlers which were generated by whistler anisotropy instability would in fact exhibit decreasing trend between the width and electron energy. The same behaviour was observed by \citet{Feldman1978,Pilipp1987a} \& \citet{Fitzenreiter1998}. Positive correlation between strahl width and electron energy was reported by \citet{Pagel2007} in the study of cases with especially broad strahl observed at 1 au by ACE spacecraft. This trend can result from scattering by whistler waves generated by $k^{-3}$ power spectrum \citep{Saito2007}. We mention two examples of the strahl scattering mechanisms that can be related to the variation of the strahl pitch-angle width with electron energy, but more mechanisms have been proposed so far. These include firehose instability generated fluctuations \citep{Hellinger2014a}, Langmuir waves \citep{Pavan2013}, lower hybrid waves \citep{Shevchenko2010}, oblique kinetic Alfv\'{e}n waves \citep{Gurgiolo2012, Chen2013}, and Coulomb collisions \citep{horaites2018d}, and are discussed latter in the article.

With an exception of analysis by \citet{Stverak2009}, none of the observational studies present the radial evolution of strahl electrons within 1 au, separated by the solar wind type. As discussed above, the strahl population is more pronounced in the fast solar wind and close to the Sun, thus it is important to study strahl properties exploring the data set from Helios missions still providing the closest in-situ measurements from the Sun. 

The two almost identical Helios spacecrafts were launched in the 70's with a mission to explore the inner most parts of interplanetary space \citep{Porsche1981}. During 10 years of active mission for the 1st spacecraft, and 3 years for the 2nd one an intriguing and currently still unique data set was produced, sampling the solar wind in the ecliptic plane with the closest perihelion of 0.29 au (Helios 2). 

In this work we provide a statistical analysis of these data with a focus on strahl electrons behaviour within 1 au. Our results in general agree with previously published works, but give us an additional insight into regions closer to the Sun, from where we were able to estimate strahl properties that will be observed during the first perihelion of the Parker Solar Probe, 0.16 au from the Sun.  

\section{Instrument Description}
\label{sec:ins}

\begin{figure*}
\centering
\includegraphics[width=\hsize]{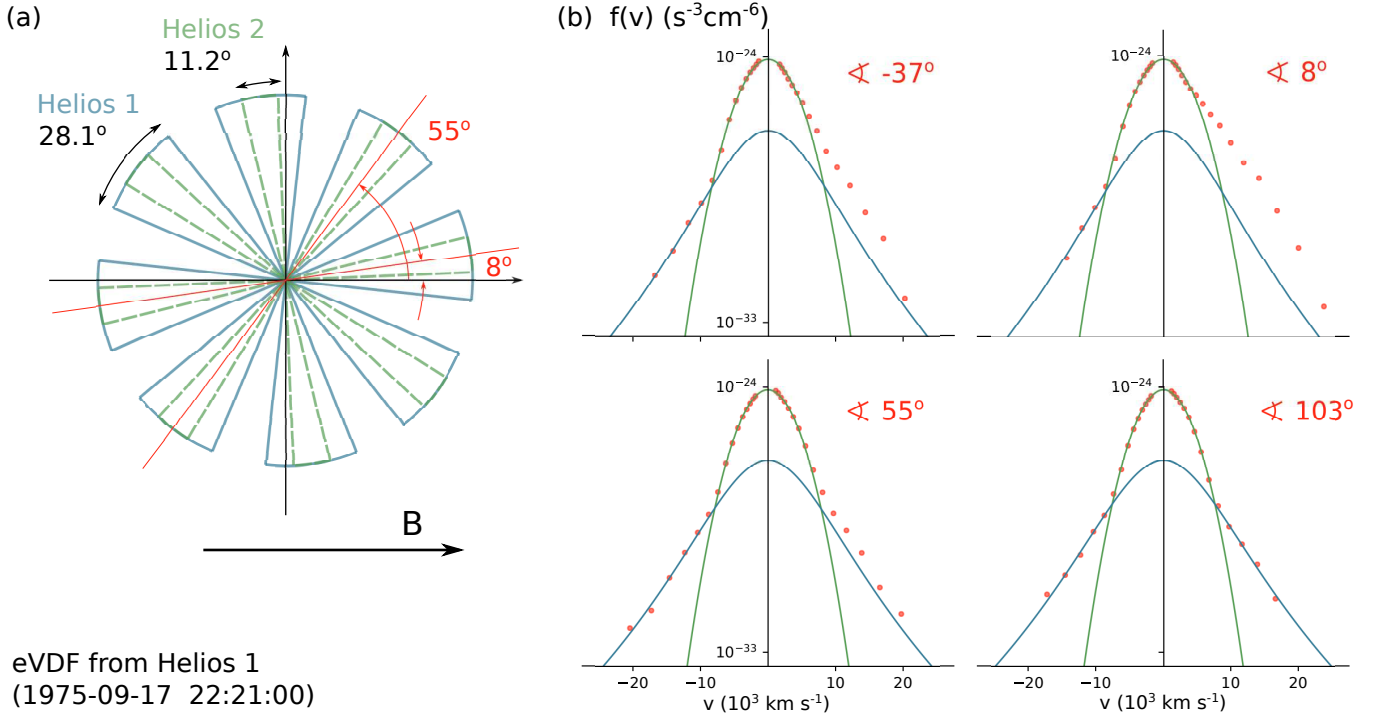}
\caption{(a) A schematics of I2 instrument azimuth sectors in the magnetic field frame corresponding to the example electron VDF shown in panel (b). The difference in sector size between Helios 1, and 2 is marked with colour. Note, however, that the example measurement was taken by Helios 1 spacecraft, and Helios 2 azimuth sectors are added to the schematics only to highlight the differences between the two. (b) An example electron VDF measured at the distance of 0.32 au from the Sun. Each of the 4 plots shows a pair of oppositely directed azimuth sectors: the red dots are measurements corrected for spacecraft potential, and green and blue line represent the fit to core and halo population, respectively. For each sector pair the angle indicates the position with respect to the magnetic field direction.}
\label{fig:ins}%
\end{figure*}

To study kinetic properties of solar wind electrons we analysed the data from the electron particle instrument I2, part of E1 Plasma Experiment on-board Helios 1 and 2 missions \citep{Rosenbauer1981,Pilipp1987}.

I2 is designed to measure a 2D distribution function of solar wind electrons within 1 au from the Sun. The instrument aperture pointing perpendicular to the spin axis of the spacecraft is followed by deflection plates, preventing sunlight-beam electrons to enter the analyser part. Electron energy is measured by a hemispherical electrostatic analyser in 16 exponentially spaced energy steps. Two different operation modes allow the measurement of either low (0.5 to 13.3 eV) or high (9 to 1445 eV) energy electrons. A channeltron sits at the end point of the electrostatic analyser and provides the electron count rate. 

The narrow instrument field of view covers 19$^o$ x 2$^o$ (elevation x azimuth) and is centred on the ecliptic plane. Both spacecrafts spin around the axis perpendicular to that plane with a spin period of 1 s allowing the instrument to sample a full 360$^o$ azimuthal angle. This is done in 8 steps (8 azimuth sectors), each lasting for 78.06 ms for Helios 1 and 31.1 ms for Helios 2, corresponding to angular sector width of 28.1$^o$ and 11.2$^o$, respectively. Thus one scan over 16 energy steps and 8 azimuthal directions is normally obtained in 16 s and repeated every 40 s.

\section{Method}

\subsection{Data set}
This study is based on the data provided by plasma experiments on-board Helios missions: the electron VDFs -- instrument I2 (described in Sec. \ref{sec:ins}), proton plasma moments -- instruments I1a and I1b, and magnetic field vectors -- instruments E2 and E3. 

The core of this analysis are electron VDFs described in section \ref{sec:evdf}. 

The proton on-board integrated densities and velocity vectors were taken from the original Helios files in Helios data archive\footnote{Link to the data archive: \emph{http://helios-data.ssl.berkeley.edu}}. The measured proton densities are likely to be underestimated, therefore the measurement with the higher value between the two -- I1a and I1b -- with 10 percent uncertainty was considered. 

The proton core temperatures we use are taken from a new Helios proton data set provided with descriptions by \citet{Stansby2018}.

Instruments E2 and E3 are the two fluxgate magnetometers on-board Helios missions. E2 samples data with a frequency of 4 Hz which is saturated at 50 nT, and E3 gives a 6s-averaged measurements. The E2 data is used if available, and if the absolute magnetic field value is smaller than 50 nT. In other cases the E3 data is used. A mean value of magnetic field vector is obtained for each 16-s electron VDF. We note that magnetic field vectors obtained this way differ from the ones used in all previous Helios data electrons studies, \textit{e.g.} \citep{Stverak2009}.   

The data set has many limitations, but we have the benefit of using measurements collected over several years by two almost identical spacecrafts. Moreover, this is the only data set providing insight on the solar wind plasma parameters in the near-Sun regions. The analysed period spans between 1974 and 1982 for Helios 1 and between 1976 and 1979 for Helios 2. We only use scans when all the above parameters are available and when the measured magnetic field vector lies within 5$^o$ from the I2 measuring plane (the ecliptic plane).

\subsection{Electron VDF}
\label{sec:evdf}

The measurements of the solar wind electrons are strongly polluted by two phenomena: photoelectrons emitted from the spacecraft body, and spacecraft charging. A method for correcting these effects making use of other in-situ plasma measurements is well described by \citet{Salem2001}.

Photoelectrons appear as a sharp peak at low energies and have already been removed in the provided Helios data set. 

A charged spacecraft deforms electron VDF depending on the shape and magnitude of the spacecraft potential which varies as a function of the surrounding plasma \citep{Pedersen2008}. In the solar wind at 1 au the typical values of spacecraft potential are between 1 and 10 V \citep{Salem2001}, and decreasing with distance from the Sun. A positive charge accelerates electrons toward the instrument making their energies seemingly larger. The density obtained by integration of this deformed VDF would therefore be overestimated.  

\citet{Salem2001} suggest the use of electron density obtained by a thermal noise receiver measuring the plasma peak to scale the VDF preforming a linear shift in electron energy. We apply the same method to determine the spacecraft potential, however, since there were no thermal noise receiver measurements made by the two Helios missions, we use a less reliable proton density measurement from I1a and I1b instruments instead. We assume quasi neutrality ($n_e = n_p + 2n_\alpha$), and a typical alpha particle to proton number ratio of 0.05.

The corrected VDF is then shifted to the plasma zero velocity frame using the proton velocity measurement, and rotated to the magnetic field aligned frame defined by the magnetic field measurement during each scan. In this frame the 0 degrees angle is aligned with the direction of either positive or negative magnetic field vector and always pointing antisunward. An example of a VDF at this point is shown in Fig. \ref{fig:ins}(b), where each of the four plots consists of two oppositely located azimuth sectors. The angles indicate how far each sector pair lays from the magnetic field direction. The sign of the angle (within the interval (-180$^o$, 180$^o$)) is kept for easier understanding of the schematics in Fig. \ref{fig:ins}(a), but it is not relevant for our further analysis.

A non-linear least squares method is used to fit two solar wind electron components: a core and a halo (see Fig. \ref{fig:ins}(b)). We do not fit strahl component because our aim is to study the energy dependent radial evolution of it, neither the superhalo component as it is out of the measuring energy range of the instrument. To model the core we use a 2D bi-Maxwellian function $f_c (v_\perp, v_\parallel)$ (see Eq. 1), and for the halo a 2D bi-Kappa function $f_h (v_\perp, v_\parallel)$ (see Eq. 2), the same model as used by \citet{Maksimovic2005}:

\begin{equation}
\begin{aligned}
A_c &=& n_c \Big(\frac{m_e}{2 \pi k_B}\Big)^{3/2} \frac{1}{T_{c\perp} \sqrt{T_{c\parallel}} } \\
f_c(v_\perp, v_\parallel) &=& A_c \exp{\Big( {-\frac{m_e}{2k_B}\Big( \frac{(v_\perp-\Delta v_{c\perp })^2}{T_{c\perp}} +\frac{(v_\parallel-\Delta v_{c\parallel})^2}{T_{c\parallel}}\Big)}\Big)}
\end{aligned}
\end{equation}

\begin{equation}
\begin{aligned}
A_h &=& n_h\Big(\frac{m_e}{\pi k_B (2\kappa-3)}\Big)^{3/2}\frac{1}{T_{h\perp} \sqrt{T_{h\parallel}}} \frac{\Gamma (\kappa+1)}{\Gamma(\kappa-\frac{1}{2})} \\
f_h(v_\perp, v_\parallel) &=& A_h \Big(1+ \frac{m_e}{k_B(2\kappa-3)}\Big(\frac{(v_\perp-\Delta v_{c\perp})^2}{T_{h\perp}} \\ &+& \frac{(v_\parallel-\Delta v_{c\parallel})^2}{T_{h\parallel}}\Big)\Big)^{-\kappa-1}
\end{aligned}
\end{equation}

In the above equations $v_{\perp}$ and $v_{\parallel}$ are the independent variables of functions $f_c$ and $f_h$. With $m_e$ we mark the mass of an electron, and with $k_B$ the Boltzman constant. Quantities $n$, $v$, and $T$ with indices $c$ -- core and $h$ -- halo, stand for the density, the velocity, and the temperature of the respective electron components. The drift velocity between the core and the halo is assumed to be 0, thus the values $\Delta v_{c\perp}$ and $\Delta v_{c\parallel}$ in Eq. 2 are the values obtained from the fit to the core population. We are left with 9 fitting parameters: $n_c$, $\Delta v_{c\perp}$, $\Delta v_{c\parallel}$, $T_{c\perp}$, $T_{c\parallel}$, $n_h$,  $T_{h\perp}$, $T_{h\parallel}$, and $\kappa$. 

To isolate the strahl population the fit ($f_c$ + $f_h$) is subtracted from the measured values. If the residual is a positive value higher than 0.9 $\cdot$ ($f_c$ + $f_h$) it is kept as a strahl VDF. The ratio 0.9 was chosen because it appears to correctly separate the core electron fit errors from the lowest strahl electron energies. We believe that a strahl component was detected, if the strahl VDF consists of at least 5 data points. In the opposite case we mark that the strahl was not observed.

\begin{figure}
\centering
\includegraphics[width=\hsize]{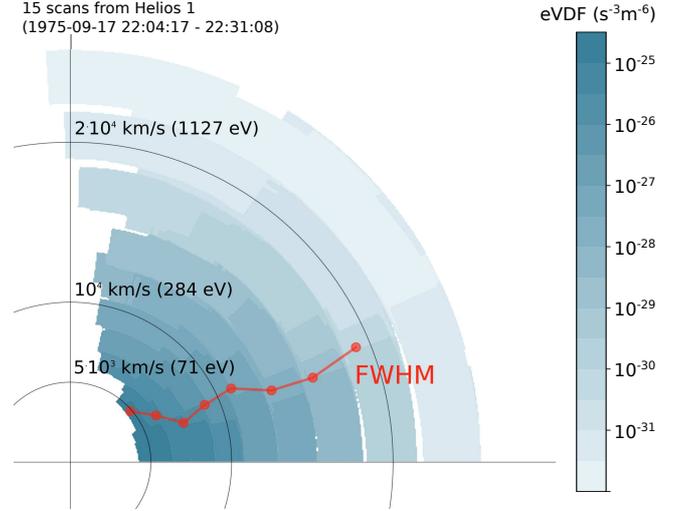}
\caption{Strahl VDF in velocity space where x-axis presents velocity parallel to the magnetic field, and y-axis the perpendicular one. The instrumental properties like azimuth sector width and energy bin size are still distinguishable.}
\label{fig:strahl}%
\end{figure}

We assume that the strahl is symmetric with respect to the magnetic field vector. As already said, these electrons are aligned with the magnetic field in the anti-sunward direction, so they can be detected by maximum 4 azimuth sectors, but commonly by only 2 of them. We enhance the angular resolution by averaging over consecutive scans, assuming that during the averaging time solar wind conditions do not change significantly. To make sure of that we only group up to 15 scans which belong to the same solar wind type (see Sec. 3.3) and satisfy the following arbitrary conditions: $\Delta$v$_p$ < 40 km/s, $\Delta n_p$ < 15 cm$^{-3}$, $\Delta$B < 10 nT, and $\Delta \Psi_B$ < 30$^o$, where $\Delta$ stands for the difference between 2 consecutive scans following the equation: $\Delta X = |{X_i - X_{i+1}}|$. Index p stands for proton, $\Delta B$ is the variation of magnetic field amplitude, and $\Delta \Psi_B$ the variation of the magnetic field angle in the ecliptic plane. Fig. \ref{fig:strahl} shows an example result of this kind of averaging in velocity space. 

In the example strahl VDF from the Helios 1 spacecraft (Fig. \ref{fig:strahl}) we can still recognise the instrumental properties: the size of the azimuth sectors (28.1$^o$) and energy bins. Even though the resolution is improved by averaging consecutive scans (with slightly different magnetic field vector direction), the smallest measurable angle stays fundamentally limited by the angular breadth of the azimuth sectors of I2 instruments (Helios 1: 28.1$^o$, and Helios 2: 11.2$^o$).

We study the width of strahl VDF, and a way to define it is using the full width half maximum parameter (FWHM), also used by e.g. \citet{Hammond1996, Graham2017}. We measure FWHM for each energy bin, by fitting the values of this bin with a normal distribution function, centred at angle 0$^o$ -- the magnetic field direction:
\begin{equation}
f(\text{PA}) = a \exp{\Big(-\frac{1}{2}\Big(\frac{\text{PA}^2}{\sigma^2}\Big)\Big)}, 
\end{equation}

where $a$, and $\sigma$ are the fitting parameters and PA stands for pitch-angle, the angle from the magnetic field direction (see Fig. \ref{fig:strahl}). This angle is defined in terms of parallel and perpendicular velocity as: 
\begin{equation}
\text{PA} = \tan^{-1}(v_\perp/v_\parallel) 
\end{equation}
FWHM is calculated from $\sigma$ parameter using $\text{FWHM} = 2\sqrt{2 \ln{2}}\cdot \sigma$, and is referred to as strahl pitch-angle width (PAW). 

\subsection{Binning}

\begin{figure*}
\centering
\includegraphics[width=1\hsize]{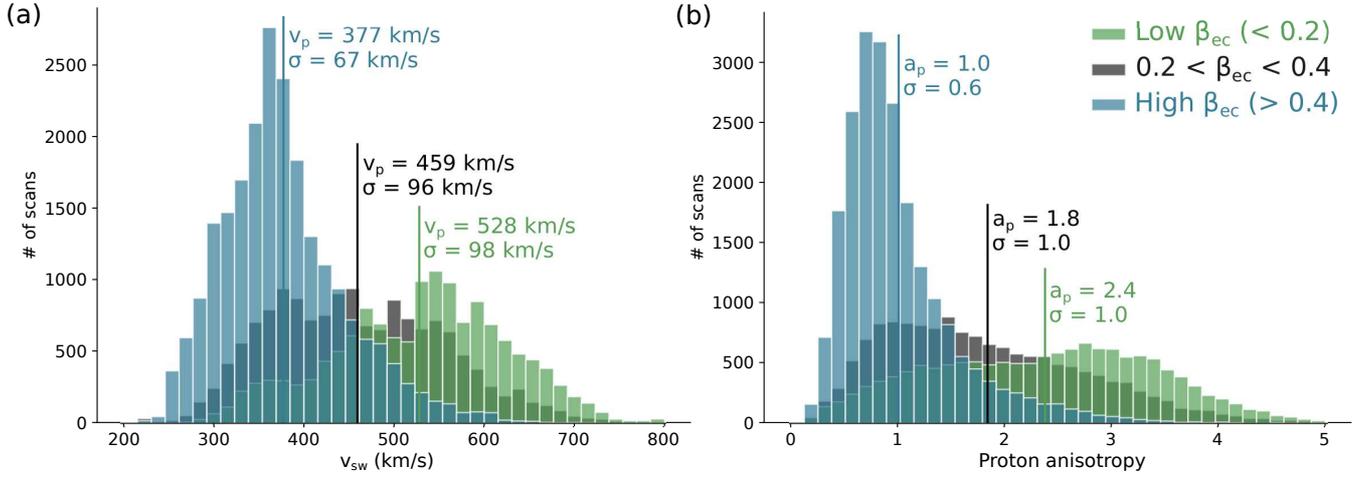} 
\caption{Histograms showing how the three solar wind types according to $\beta_{ec\parallel}$ relate to the solar wind velocity -- (a), and proton core anisotropy -- (b). The mean values with the standard deviations for each type are marked in both plots.}%
\label{fig:fast-slow}
\end{figure*}

The solar wind is usually separated into fast, and slow wind according to its proton velocity. Another interesting separation was proposed in a recent work of Stansby et al., 2018, where the solar wind is separated into 3 types: slow alfvenic, slow non-alfvenic, and fast alfvenic wind, by its measured proton anisotropy and cross helicity. Even though both of the mentioned separation techniques give the same main observational results of the present article for the fast, and the slow solar wind, we find that it is better to separate the solar wind into types according to a parameter more closely related to the kinetic properties of the solar wind electrons. In the following sections the solar wind is separated according to core electron parallel beta value ($\beta_{ec\parallel}$), the ratio of plasma parallel pressure to magnetic pressure, defined as:

\begin{equation}
\beta_{ec\parallel} = \frac{2 \mu_0 n_c k_B T_{c \parallel}}{B^2}, 
\label{eq:ecpar}
\end{equation}

where $\mu_0$ is the permeability of free space, and $B$ the magnitude of the measured magnetic field. We chose $\beta_{ec\parallel}$ as a separation parameter because it spans over a large range of more than two magnitudes, but does not exhibit a radial dependency. This is not true for the halo electron parallel beta ($\beta_{eh\parallel}$), which is observed to increase with the radial distance (see Fig. \ref{fig:ani-beta}(b)). We define three solar wind types: low-$\beta_{ec\parallel}$ wind ($\beta_{ec\parallel}$ < 0.2), intermediate-$\beta_{ec\parallel}$ wind (0.2 < $\beta_{ec\parallel}$ < 0.4), and high-$\beta_{ec\parallel}$ wind ($\beta_{ec\parallel}$ > 0.4). The arbitrary chosen separation values are marked in an electron anisotropy-$\beta_{ec\parallel}$ parameter space in Fig. \ref{fig:ani-beta}(a) with red dashed lines.

How our solar wind separation compares to the solar wind proton velocity and anisotropy \citep{Matteini2007} is shown with histograms in Fig. \ref{fig:fast-slow}. The low-$\beta_{ec\parallel}$ wind corresponds to the faster solar wind with higher proton anisotropy averaging to $a_p = 2.4$, while the high-$\beta_{ec\parallel}$ wind represents the slow almost isotropic solar wind.

The data set is naturally binned in energy by instrumental energy bins, and additionally according to the distance from the Sun into 7 equally spaced bins. A mean value of the strahl pitch angle width with its standard error is assigned to each bin. 

Starting from 231,778 scans with the magnetic field vector close to the ecliptic plane, 51,570 were successfully fitted with models for core and halo components and matched with the solar wind proton data. Of these 14,052 (27\%) were identified as the low, 15,060 (29\%) as the intermediate, and 22,263 (44\%) as the high-$\beta_{ec\parallel}$ solar wind. The mean velocity of the low-$\beta_{ec\parallel}$ wind is 528 km/s, the intermediate 459 km/s, and the high-$\beta_{ec\parallel}$ wind 377 km/s. Strahl was not observed in 4,359 examples. This means that strahl was absent in 8.5\% of our observations with a mean velocity, and a standard deviation of 441, and 105 km/s. This is much less than $\sim$20\% observed by \citet{Gurgiolo2017} or 25\% by \citet{Anderson2012}. This difference might be due to the fact that most of our measurements were taken within 1 au, while both of the mentioned studies are based on the analysis of the  data from 1 au, which is consistent with the gradual disappearance of strahl with radial distance \citep{Maksimovic2005, Stverak2009, Graham2017}.

It is important to note that the number of the fast solar wind samples is decreasing with radial distance. This is because our data set, and analysis are limited for low plasma densities. The proton measurement is less accurate for low proton densities, therefore making our estimation of the spacecraft potential more inaccurate, which deforms the electron VDF and results in an unsuccessful fit.  

Another instrumental limitation could be the time needed to obtain one 2D VDF scan. We checked how much the magnetic field angle varies during the sampling time (16 s), and found no correlation between broader strahls and the variation of magnetic field angle. The standard deviation varies between 1.5 and 5.5$^o$, where larger values were found in the low-$\beta_{ec\parallel}$ solar wind.

\section{Observations}

\begin{figure}
\centering
\includegraphics[width=1\hsize]{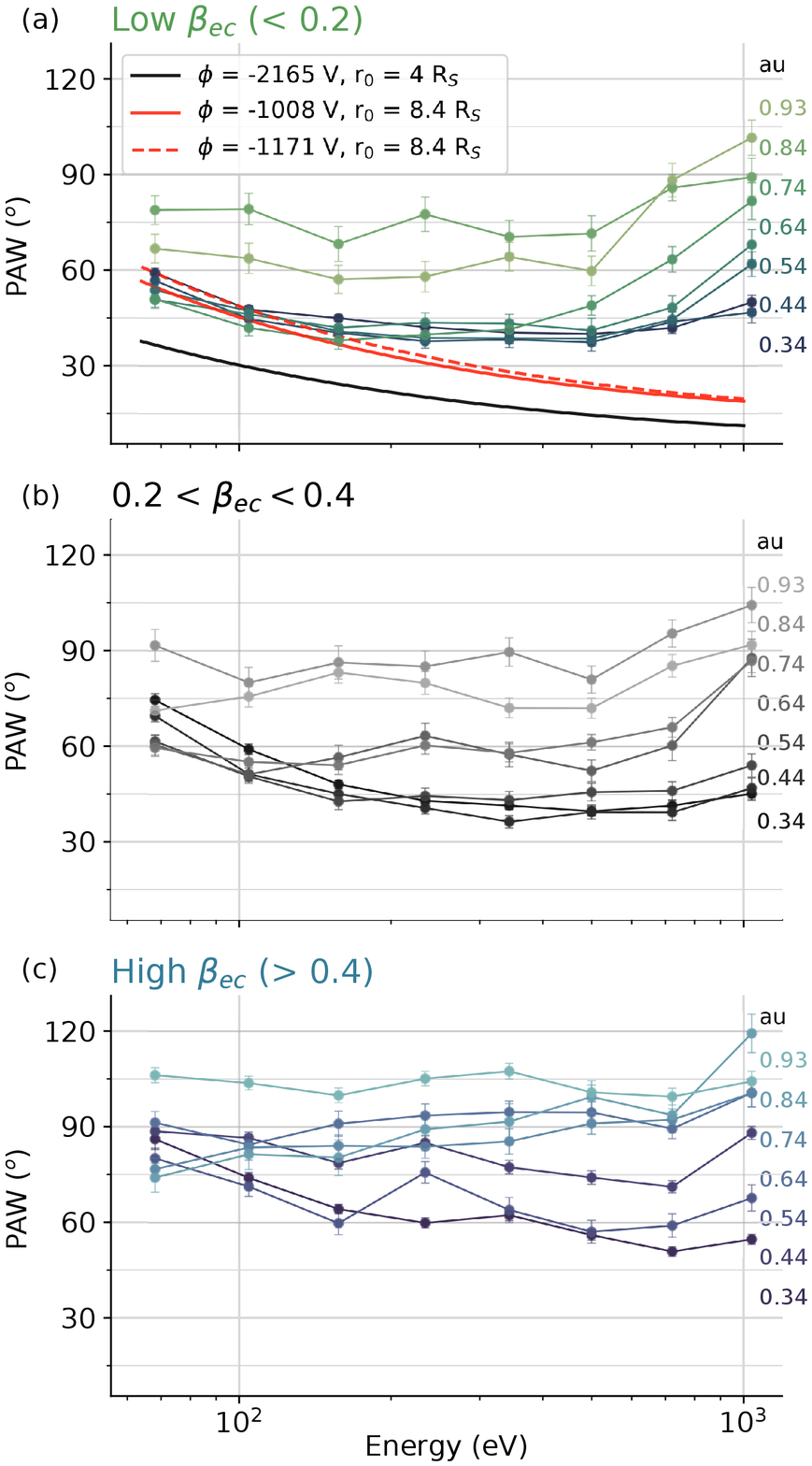}
\caption{Strahl pitch-angle width versus electron energy shown separately for the low -- (a), intermediate -- (b), and high -- (c) $\beta_{ec\parallel}$ solar wind. Darker coloured lines denote distances closer, and lighter coloured lines distances farther from the Sun. In the upper plot a dashed red, a solid red, and a solid black line denote a curve resulting from a simple collisionless focusing model for three different parameter pairs (see Sec. \ref{lowbeta}).}%
\label{fig:pa-en}
\end{figure}

Different properties of strahl electrons were found for the low-, intermediate-, and the high-$\beta_{ec\parallel}$ wind. For each of them, Fig. \ref{fig:pa-en} shows how strahl PAW varies with electron energy. Differently coloured lines represent different distances from the Sun. We focus on the low-, and the high-$\beta_{ec\parallel}$ type, as the intermediate possesses the properties of both of them. 

The strahl component in the low-$\beta_{ec\parallel}$ wind, that can be related to the fast solar wind, appears narrower than in the high-$\beta_{ec\parallel}$ case. The PAW properties depend on the electron energy. For the lower energies, up to 343 eV the PAW is decreasing with electron energy. The PAWs vary very little between 0.34 to 0.74 au, however, a slight decrease with radial distance is observed in this low electron energy range. The PAW seems to saturate just below 40$^o$, which is an effect of a limited angular resolution of the electron instrument I2.

Interestingly, for the electron energies above the 499 eV bin strahl PAW increases with electron energy and the distance from the Sun.

Strahl electrons in the high-$\beta_{ec\parallel}$ wind appear more than 20$^o$ wider than in the low-$\beta_{ec\parallel}$ wind already at 0.34 au. An anti correlation between PAW and electron energy can be observed. Moving away from the Sun, the strahl is becoming broader and less correlated with electron energy. At the distance of 0.94 au from the Sun PAW is no longer correlated with the energy and reaches values above 100$^o$.

\section{Discussion}

\subsection{Low-$\beta_{ec\parallel}$ solar wind}
\label{lowbeta}

We observe that the strahl electrons in the low-$\beta_{ec\parallel}$ solar wind exhibit different trends depending on their energy. 
In the low energy part observations of strahl electrons for the first time show a slight decrease in the strahl PAW with the radial distance. All the existing observational studies of the evolution of strahl PAW with distance \citep{Hammond1996, Graham2017, Graham2018} show a broadening of the strahl during expansion, however, none of them samples the radial distances below 0.8 au, where the focusing was found in the present work. Thus, the decrease of PAW with distance is particular for the low-$\beta_{ec\parallel}$ solar wind, and for the regions closer to the Sun (down to 0.3 au).

As mentioned in the introduction, the strahl electrons are the electrons which at some distance close to the Sun escape the dense corona dominated by collisions, and during their escape undergo the focusing effect induced by the radially decreasing magnetic field. We present a simple collisionless focusing model, often used in the exospheric models (\textit{e.g.} \citet{Maksimovic1997}), to understand what would be the shape of the electron strahl originating from an isotropic function close to the Sun at the point of our first observations, at 0.34 au. We assume that at a given point an isotropic distribution function starts to focus conserving the electron energy and the magnetic moment:

\begin{equation}
\frac{m_e}{2}(v_\perp^2 + v_\parallel^2)  - e \Phi = \text{const.} \ \ \ \ \ \ \text{and} \ \ \ \ \ \ \ \frac{m_e v_\perp^2}{2 B} = \text{const.}\
\label{eq:f1}
\end{equation}

In equations $e$ stands for elementary charge, and $\Phi$ for the electrostatic ambipolar potential in the solar wind, with $\Phi = 0$ at infinity. We now write these equations indexing quantities at the focusing starting point with 0, and at the distance of our first observation (0.34 au) with 1. The strahl PAW of the isotropic distribution function at the focusing starting point is described with the PAW of 180$^o$, thus the parallel velocity ($v_{\parallel 0}$) equal to 0.  

\begin{equation}
\frac{m_e}{2}v_{\perp 0}^2 - e \Phi_0 = \frac{m_e}{2}v_{\parallel 1}^2 + \frac{m_e}{2}v_{\perp 1}^2 - e \Phi_1, 
\label{eq:f3}
\end{equation}

\begin{equation}
\frac{m_e v_{\perp 0}^2}{2 B_0} = \frac{m_e v_{\perp 1}^2}{2 B_1}.
\label{eq:f4}
\end{equation}

From Eq. \ref{eq:f3} and \ref{eq:f4} we obtain expressions for parallel, and perpendicular velocities at the observation point,

\begin{equation}
v_{\parallel 1}^2 = v_{\perp 0}^2 \Big(1 - \frac{B_1}{B_0} \Big) + \frac{2e}{m_e} \Delta\Phi
\label{eq:f5}
\end{equation}

\begin{equation}
v_{\perp 1}^2 = v_{\perp 0}^2 \frac{B_1}{B_0},
\label{eq:f6}
\end{equation}

where $\Delta \Phi = \Phi_1 - \Phi_0$ is the difference in electrical potential between the observation and the starting point. To compare the model directly to our observations in Fig. \ref{fig:pa-en} we would like to express the model PAW in terms of electron energy (E):
\begin{equation}
E = \frac{m_e}{2}v_{\perp 0}^2 + e\Delta \Phi
\label{eq:f7}
\end{equation}

Using eq. \ref{eq:f7} we can rewrite the expressions for the parallel and the perpendicular velocity (eq. \ref{eq:f5} and \ref{eq:f6}) as:

\begin{equation}
\frac{m_e}{2} v^2_{\parallel 1} = E - \frac{B_1}{B_0}\Big( E -e\Delta \Phi\Big)
\label{eq:f8}
\end{equation}

\begin{equation}
\frac{m_e}{2} v^2_{\perp1} = \frac{B_1}{B_0}\Big( E -e\Delta \Phi\Big) 
\label{eq:f9}
\end{equation}

Combining eq. \ref{eq:f8} and \ref{eq:f9}, we obtain an expression for the strahl PAW of the model distribution at $r_1$, which we denote as PAW$_{cf}(E)$:

\begin{equation}
\text{PAW}_{cf} (E) =  2\cdot\tan^{-1}{\Big( \frac{v_{\perp 1}}{v_{\parallel 1}}\Big)} = 2\cdot \tan^{-1}{\Big(\sqrt{ \frac{E - e\Delta \Phi}{\frac{B_0}{B_1}E - E + e\Delta \Phi}}\Big)}
\label{eq:f10}
\end{equation}

We find that PAW$_{cf}$ is a decreasing function of energy if the magnetic field strength, and the electric potential are decreasing with the distance from the Sun. This is normally true in the solar wind. We calculate the PAW$_{cf}$ for a simplified case where we assume that magnetic field strength changes with $r^2$ and use the electric potential values from a transonic collisionless model of the solar wind by \citet{Zouganelis2004}\footnote{The electrostatic potential values are taken from Fig. 1 for $\kappa$ = 2.5, $r_0$ = 4 $R_S$: $\Delta \Phi = -2165 V$}. The value $r_0$, the focusing starting point, is set to 4 solar radii, following \citet{Maksimovic1997b}, who find that in their kinetic model of the solar wind with Kappa distribution functions the exobase is located between 2.8 and 10.2 solar radii, where the distance 4 solar radii corresponds to typical equatorial region solar wind conditions.
This solution shown with a black line in Fig. \ref{fig:pa-en}(a) gives a strahl component which is about half the width of the strahl observed for low energies at 0.34 au. 

Still assuming that the magnetic field strength decreases with $r^2$, we fit the model to the PAWs observed for the lowest two energies at 0.34 au (the dashed red line in Fig. \ref{fig:pa-en}(a)). To recover the observed strahl width the focusing of the solar wind electrons needs to start further away from the Sun, at the distance of $r_0$ = 8.4 $R_S$, which is still in the range discussed by \citet{Maksimovic1997b}. The potential difference obtained from the fit ($\Delta \Phi$ = -1171 V) is very close to the one taken for the same $r_0$ from the model of \citet{Zouganelis2004}\footnote{For $\kappa$ = 2.5, $r_0$ = 8.4 $R_S$: $\Delta \Phi = -1008 V$}. For comparison the strahl PAW solution according to the model of \citet{Zouganelis2004} for $r_0$ = 8.4 is plotted in Fig. \ref{fig:pa-en}(a) with a red solid line.  

We conclude that the strahl PAWs observed for the low electron energies close to the Sun could be a result of collisionless focusing of the solar wind electrons during expansion. The shape of the observed strahl distribution function at 0.34 au corresponds well to the shape predicted by a collisionless focusing model with parameters in the range of the ones reported for the solar wind.\\

Even though a slight focusing over radial distance is observed for the low energies of the low-$\beta_{ec\parallel}$, the PAW decrease is not strong enough to follow collisionless focusing described by Eq.\ref{eq:f1}. 
We consider collisions as a possible strahl scattering mechanism in this low strahl electron energy range. In the future we plan to make use of a fully kinetic solar wind simulation \citep{Landi2012, Landi2014} to explore the limiting energy at which the Coulomb collisions are still able to effect the electron VDF. However, the lowest strahl energy presented in this article, 68 eV, already equals to more than 3 times the typical core electron thermal energy, so collisions are expected to be very rare.\\

The positive correlation between strahl PAW and electron energy, observed for the more energetic strahl electrons in the low-$\beta_{ec\parallel}$ solar wind, was already reported in the study of \citet{Pagel2007}. The authors analyse 29 events during times when extremely broad strahl was observed at 1 au. The mean solar wind velocity of these 29 events, 501 km/s, is comparable to the mean velocity of our low-$\beta_{ec\parallel}$ population, 528 km/s. From the relation between PAW and electron energy they conclude that the source of the scattering of the strahl electrons are most likely the quasi-parallel broad-band whistler-mode waves generated by the magnetic field power spectrum in the whistler range. The cyclotron resonance of the faster electrons corresponds to smaller k-vectors, for which the magnetic field fluctuations are larger in the solar wind, providing stronger scattering of the higher energy electrons.

Supporting this hypothesis are the particle-in-cell simulations provided by \citet{Saito2007}, and a kinetic model in a framework of quasi-linear theory by \citet{Vocks2005}. However, sunward directed wave k-vectors parallel to the background magnetic field needed for whistlers to be able to resonate with anti-sunward moving electrons \citep{gary_1993} were observed to be rare at sub-ion scales. Moreover, \citet{Chen2010} observe the power in the parallel spectral component ($\delta B(k_\parallel)^2$) to be only 5 \% of the power in perpendicular one ($\delta B(k_\perp)^2$). Another possibility is that the correlation between the strongly scattered faster strahl electrons and the magnetic field power spectrum results from a mechanism related to the perpendicular magnetic field fluctuations. An example of this kind of mechanism is stochastic heating studied for the case of solar wind protons by \citet{Chandran2013}. To our knowledge no similar theory has been developed for electrons so far. 

\begin{figure*}
\centering
\includegraphics[width=1\hsize]{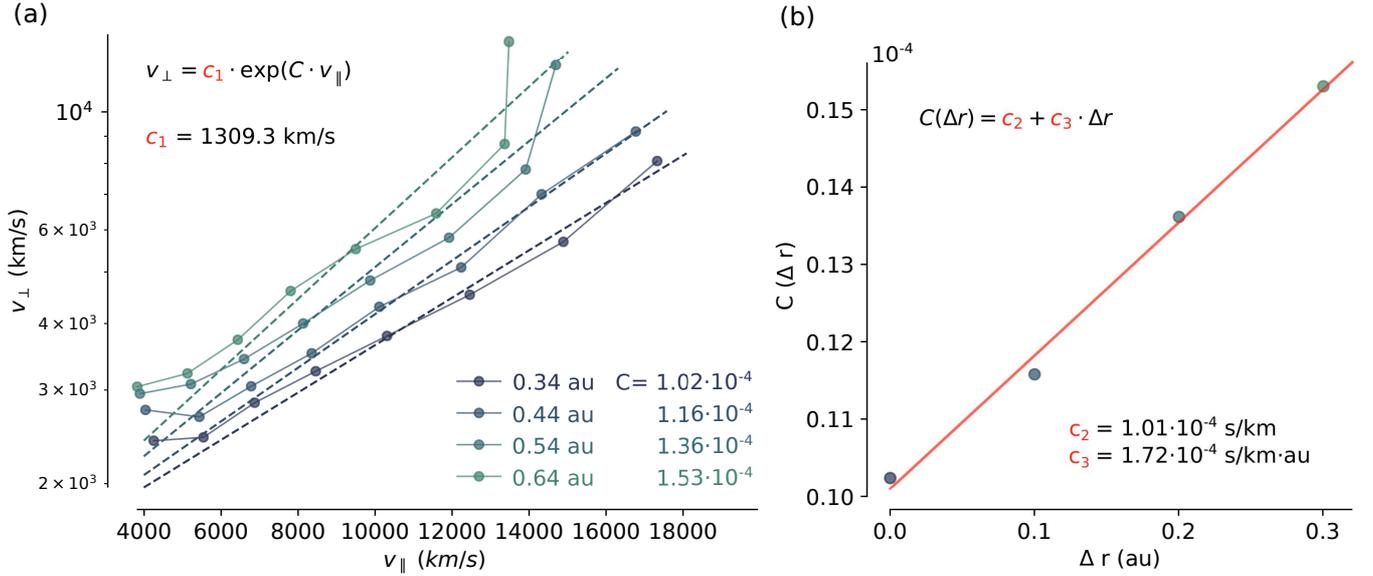} 
\caption{Comparison between the empirical model (dashed line), and observations for 4 closest distances from the Sun of low-$\beta_{ec\parallel}$ solar wind (dots). }%
\label{fig:model}
\end{figure*}

Variations in the magnetic field could affect the trajectories of the gyrating electrons if their gyroradius would be of the same scale as the changes in the magnetic field. Typically the gyroradius of the strahl electrons, directly proportional to their perpendicular velocity, spans between a few tenths and 30 km, larger radii corresponding to more energetic electrons. We can now again draw the correlation with the magnetic field power spectra: k-vectors are inversely proportional to the gyroradii, and the amplitude of the fluctuations in the solar wind, thus more energetic electrons are diffused by the stronger fluctuating magnetic field. It should be noted that this diffusion process has not yet been studied in detail, and is for now just a candidate to explain our observations.

To better quantify the observations presented in this article, a simple empirical model of the scattering of the strahl components is proposed. The mechanism at work has to first overcome the theoretically predicted focusing effect, and then further scatter strahl electrons. We estimate how strong the focusing is for each radial distance starting from the observed strahl at 0.34 au and applying the electron energy and magnetic moment conservation (see Eq. \ref{eq:f1}). As above, the electric potential values are taken from the work of \citet{Zouganelis2004}.
By adding to the observed strahl PAW the angle for which the strahl has been focused over a given radial distance we obtain the total-required-scattering PAW, used in Fig. \ref{fig:model}. We only consider distances from 0.34 to 0.64 au from the Sun, as at larger distances the strahl PAWs do not appear to follow a continuous function anymore (see for example most right plot in Fig. \ref{fig:vel}(a)) and the measurements become less reliable due to the higher relative error on the measurement of the solar wind density.  
We can describe these PAWs with a perpendicular scattering process, in which electrons are scattered across the magnetic field as a function of their parallel velocity ($v_\parallel$) with an empirical exponential form:

\begin{equation}
v_\perp ( v_\parallel ) = c_1 \cdot \exp{\big[ C \cdot v_\parallel}\big],
\label{eq:vperp}
\end{equation}

where $c_1$ and C are the fitting parameters. This can be easily seen in Fig. \ref{fig:model} (a), as the higher energy part of strahl PAWs observed at different radial distances form almost straight lines in linear-logarithmic scale space. We find that the first parameter, $c_1$, does not vary significantly with the radial distance and can be fixed to a value 1309.3 km/s. The later parameter, C, depends on the distance from the Sun as shown in Fig. \ref{fig:model} (b). We can write C, and consequentially $v_\perp$ as

\begin{equation}
C (\Delta r) = c_2 + c_3 \cdot \Delta r \ \ \ \ \ and  \ \ \ \ \ \Delta r = r - 0.34 \text{au},
\label{eq:dr}
\end{equation}

\begin{equation}
v_\perp (v_\parallel, \Delta r) = c_1 \cdot \exp{\big[( c_2 + c_3 \cdot \Delta r) \cdot  v_\parallel \big]}.
\label{eq:vperp1}
\end{equation}

$r$ stands for the distance from the Sun for each of the observations.
Note that this type of scattering does not necessarily conserve the particle total energy. As electrons are scattered in perpendicular direction they can take the energy from the scattering source (i.e. ambient electromagnetic waves, or background turbulence), and if that is the case, their parallel velocity can remain unchanged.

The values of fitting parameters are noted in Table \ref{table:1}. 

An exponential relation between $v_\perp$ and $v_\parallel$ is in a case when $v_\parallel$ does not vary with distance (the total particle energy is not conserved) a solution of the differential equation which can be written as:
\begin{equation}
\frac{1}{v_\perp} \frac{dv_\perp}{dr} = c_3 \cdot v_\parallel, 
\label{eq:dif}
\end{equation}

where the constant $c_3$ describes the scattering strength. We would like to emphasise that this model is solely empirical and is developed with the purpose to better understand the observations. Further studies of the scattering mechanisms are required to understand whether a physical phenomena (or a combination of them) can result in above described velocity dependent scattering. \\

\begin{table}
\caption{Fitting parameters}            
\label{table:1}  
\centering       
\begin{tabular}{c|c|c}     
 $c_1$ & $c_2$ & $c_3$ \\     
\hline                    
 1309.3 km/s & 1.01 $\cdot$ 10$^{-4}$ s/km & 1.72 $\cdot$ 10$^{-4}$ s/(km $\cdot$ au) \\      % inserting body of the table
                          
\end{tabular}
\end{table}

\subsection{High-$\beta_{ec\parallel}$ solar wind}

For the high-$\beta_{ec\parallel}$ solar wind with the mean velocity of 377 km/s, scattering of the strahl electrons appears to be extremely efficient over the whole electron energy range. In the work of \citet{Gurgiolo2017} the authors show that 20 \% of the observed solar wind at 1 au with velocities below 425 km/s appears without the strahl electron component, and pose the question whether this is a consequence of the strahl origin, or of some transit mechanisms acting upon it during its expansion. Our radial dependent observations confirm the later: during the radial evolution, the strahl broadens until the point when it is completely scattered into the halo component. Electron velocity distribution functions without the strahl were mainly observed in the high-$\beta_{ec\parallel}$ solar wind at larger distances from the Sun.

\begin{figure*}
\centering
\includegraphics[width=0.95\hsize]{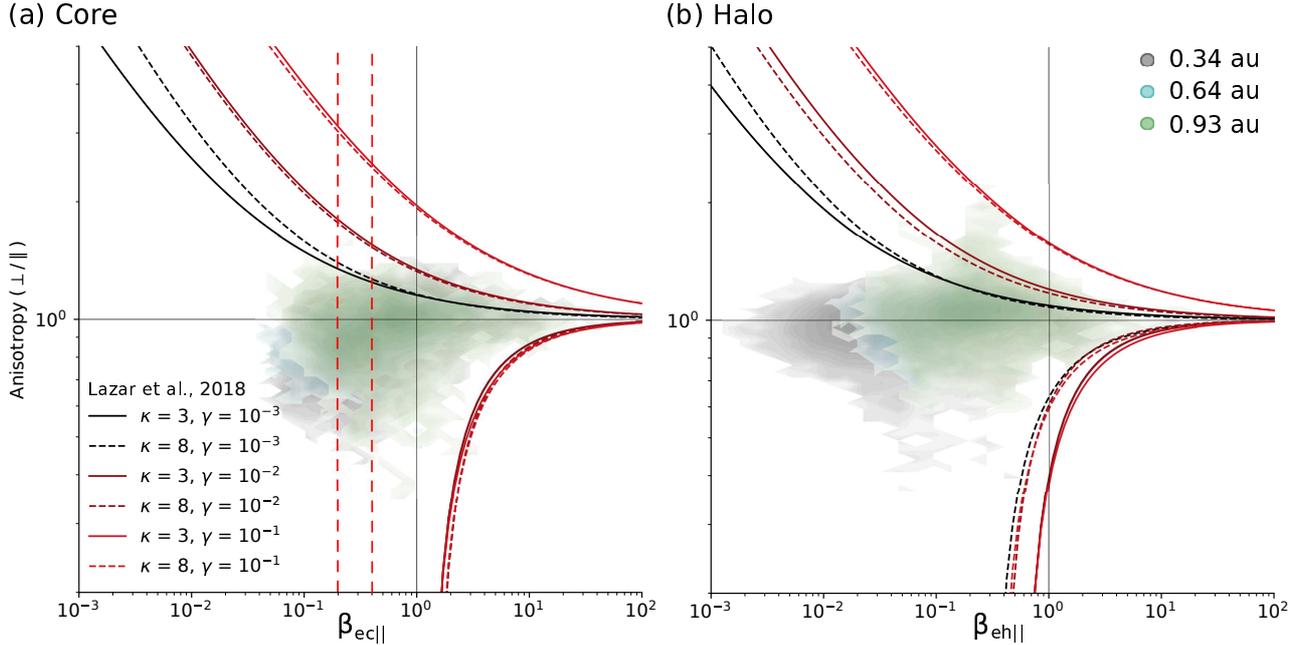} 
\caption{A contour plot showing radial evolution of core -- (a), and halo -- (b) electrons in anisotropy-beta parameter space. Three colours denote measurements taken within three different 0.1 au wide radial bins centred on the values given in the plot. The blue lines present the whistler instability ($a_e > 1$), and firehose instability ($a_e < 1$) maximum growth rate curves obtained by \citet{Lazar2018}. Two red dashed lines show the arbitrary chosen $\beta_{ec\parallel}$ values separating solar wind into three types.}%
\label{fig:ani-beta}
\end{figure*}

A reason for this efficient scattering might lay in the $\beta_{ec\parallel}$ parameter itself. This dense population of the solar wind electrons takes up a region of the $\beta_{ec\parallel}$-anisotropy parameter space constrained by instabilities, e.g. whistler, or firehose  instability (given in Fig. \ref{fig:ani-beta}). 

A direct correlation between narrow-band whistler activity and the broadening of the strahl was presented by \citet{Kajdic2016}. On the basis of statistical study of the slow solar wind (the velocity is below 400 km/s for most of the samples) at 1 au they conclude that anti-correlation between the PAW and electron energy is broken in the presence of narrow-band whistler-mode waves which scatter a portion of strahl velocity distribution function. Note that in this work the direction of the detected whistler waves could not be inferred. The broadening is energy dependent, spanning from 5$^o$ to 28$^o$ \footnote{These values were converted to FWHM for consistency. Strahl PAW in the analysis by \citet{Kajdic2016} is defined as standard deviation, $\sigma$, and $\text{FWHM} = 2\sqrt{2 \ln{2}}\cdot \sigma$.} influencing electrons with energies between 250 and 600 eV. In our data set decrease of the PAW with energy was not observed close to 1 au, but very similar tendencies were found in the slow solar wind closer to the Sun: PAW decrease with electron energy, and broader strahl for energies between 200 and 500 eV. The source of these whistles, however, is not discussed in the above cited work.

Properties of whistler-mode waves observed in near-Earth regions were studied by \cite{Lacombe2014}. Authors believe that whistlers are most likely generated by the whistler heat flux instability, as they were found at times when electron distributions were close to this instability threshold. Their observations show that electron temperature anisotropy ($T_\perp / T_\parallel$, taken as moments of a total VDF) is most of the time below unity, therefore excluding a possibility that whistlers are created by the whistler anisotropy instability. 

However, our obtained anisotropies separately for core, and halo components, plotted against the whistler and firehose instability conditions, which were calculated for an electron VDF consisting of maxwellian core, and a kappa halo \citep{Lazar2018} give the impression that both instabilities, limiting high $\beta_{ec\parallel}$ values could play a role in the generation of whistler-mode waves (see Fig. \ref{fig:ani-beta} (a) -- core, and (b) -- halo).

Whistler-mode waves generated by the heat-flux instability have a preferred propagation direction in the direction of the positive heat flux \citep{Gary1975}. Thus, this kind of waves will propagate away from the Sun, and will not be able to interact with strahl electrons. An observational study by \citet{Stansby2016} indeed shows that 98\% of the measured whistlers propagate in the anti-sunward direction. Anyhow, the generation of whistlers itself could change the shape of the strahl velocity distribution function. With a tendency towards a more stable, isotropic state, the strahl electrons' parallel velocities will decrease while their perpendicular velocities will increase. 

Symmetric whistlers, parallel and anti-parallel to the magnetic field direction, can theoretically develop from the whistler anisotropy instability (\textit{e.g.} of a symmetric halo component), and sunward directed portion of them could resonate with strahl electrons, enhancing their perpendicular velocities. A numerical simulation of this mechanism \citep{Saito2007} predicts a scattered strahl with a negative correlation between PAW and electron energy, as observed closer to the Sun in this work, and at 1 au by \citet{Kajdic2016}. However, as mentioned in the previous paragraph, this is in contradiction with the observations showing that sunward directed whistlers are extremely rare at 1 au \citep{Stansby2016}.\\

An alternative scattering source to whistler waves are self generated Langmuir waves discussed by \citet{Pavan2013}. Using numerical simulations the authors show that Langmuir waves can contribute to the widening of the strahl component resulting in an anticorrelation between PAW and energy, however, the velocities at which the diffusion is effective only reach up to 2 times the thermal speed of electrons. The directly observed scattering of the strahl electrons into the halo reported by \citet{Gurgiolo2012} appears at similar energy scales. In this last work the proposed source of scattering are the highly oblique kinetic Alfv\'{e}n fluctuations, which can widen the strahl through Landau damping. These two scattering mechanisms both take place at lower energies and could be effective up to $\sim$100 eV.

\subsection{Estimations of strahl pitch-angle width (PAW) closer to the Sun}

\begin{figure*}
\centering
\includegraphics[width=1\hsize]{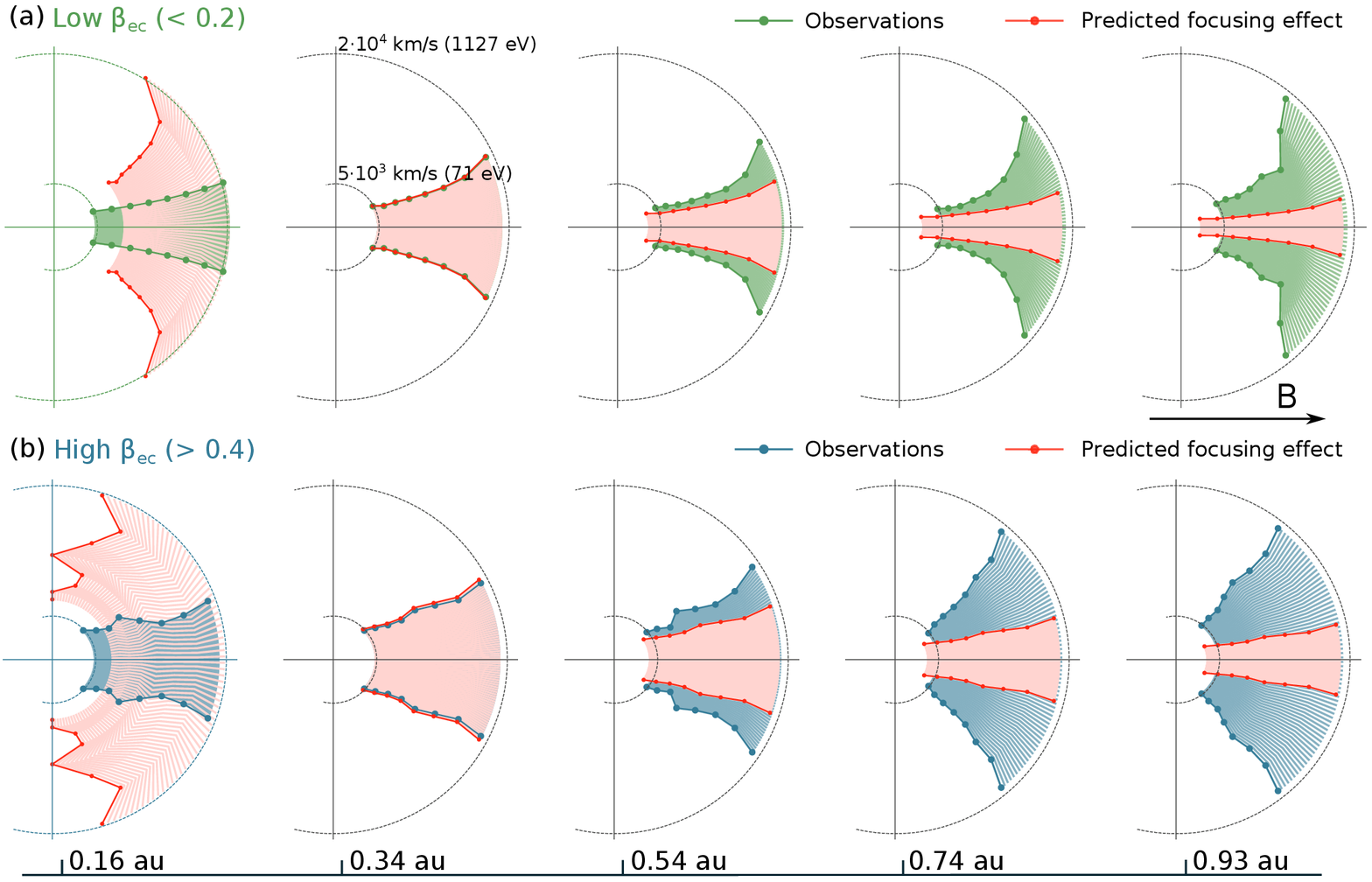}
\caption{Radial evolution of the electron strahl component in velocity space for the low -- (a), and the high -- (b) -$\beta_{ec\parallel}$ solar wind. The radial position of each plot is marked on the bottom of the figure. The left most plot, marked with a distance of 0.16 au is an estimation, in (a) obtained from the empirical model developed above, and in (b) a linear extrapolation of the observations. In red we show the shape of the strahl component resulting only from collisionless focusing. }
\label{fig:vel}%
\end{figure*}

The radial evolution of the strahl is shown in Fig. \ref{fig:vel} in velocity space, separately for each solar wind type. Green and blue colour present observations at different distances from the Sun (marked at the bottom), however the left most plots marked with a radial distance of 0.16 au are estimated from the observations. For the low-$\beta_{ec\parallel}$ solar wind an empirical relation between parallel and perpendicular strahl electron velocity (developed in the Sec. \ref{lowbeta}) was used to estimate the strahl PAW closer to the Sun, while for the high-$\beta_{ec\parallel}$ solar wind the PAW values are linearly extrapolated from the observations. Linear, the simplest, extrapolation technique is used because no model has been developed for the high-$\beta_{ec\parallel}$ solar wind.

With red colour we present how efficient is the collisionless focusing effect starting from the observation at 0.34 au. This is the same focusing model as used in Sec. \ref{lowbeta} and Fig. \ref{fig:model} taking the electrostatic potential values from the work by \citet{Zouganelis2004}.

We choose to extrapolate our observations to the distance of 0.16 au as this will be the first perihelion of the Parker Solar Probe \citep{Fox2016}. We believe that the strahl electrons will be observed narrower than at 0.34 au in the high-$\beta_{ec\parallel}$, as well as for the energies above $\sim$200 eV in the low-$\beta_{ec\parallel}$ solar wind. Using the empirical model for the low-$\beta_{ec\parallel}$ solar wind we predict that the positive correlation between the strahl PAW and electron energy will no longer be present at 0.16 au, in fact, the strahl PAW will become almost independent on the electron energy with a mean value of $\sim$ 29$^o$ (left most plot of Fig. \ref{fig:vel}(a)). Considering the  limitation of the I2 instrument in measuring small PAs (minimal angular width $\sim$28$^o$), we believe this will be the upper limit for the strahl PAW observed at 0.16 au. We expect the high-$\beta_{ec\parallel}$ solar wind strahl to be broader, between 37$^o$ and 65$^o$ (see left most plot of Fig. \ref{fig:vel}(b)).

The low energy strahl electrons (bellow $\sim$ 200 eV) in the low-$\beta_{ec\parallel}$ solar wind are observed to focus slightly during expansion already between 0.34 and 0.74 au, and we believe that the focusing effect will be observed, and even stronger at 0.16 au. The shape of the strahl component will coincide with the collisionless focusing model shown in Fig. \ref{fig:pa-en}(a). Closer to the Sun, during the upcoming perihelions, we should be able to observe the shifting point between focusing and scattering with radial distance for the higher electron energies as well.

\section{Conclusions}

An observational study of the electron strahl width in the inner Solar system reveals different behaviour of the strahl depending on the value of the electron core beta ($\beta_{ec\parallel}$) in the solar wind. 

Strahl electrons appear narrower in the low-$\beta_{ec\parallel}$ -- faster, and more tenuous -- solar wind, and their behaviour is closely related to their energy. The slower strahl electrons experience anti-correlation between PAW and their energy, and a slight focusing over radial distance for distances up to 0.74 au. Comparing the observations to a simple collisionless focusing model, we find that the strahl observed at 0.34 au for the lower energies could result from the collisionless focusing. Model parameters, $r_0$ and $\Delta \Phi$, found from fitting the data are very close to the ones reported for the solar wind.

More energetic strahl electrons show a correlation between the strahl PAW and their energy, for which we develop a simple empirical model. We observe that the increase of the electron $v_\perp$ is exponentially related to the electron $v_\parallel$ and the change in radial distance $\Delta r$. Further studies are required to understand which phenomena could scatter strahl electrons in this particular way described with Eq. \ref{eq:dif}.

Strahl electrons in the high-$\beta_{ec\parallel}$ solar wind are effectively scattered over their whole energy range. From an anti-correlation between the PAW and electron energy at 0.34 au, the strahl gets scattered to PAs above 100$^o$ close to the orbit of the Earth, many times disappearing completely from the electron VDF.  We believe that this efficient scattering is a consequence of high-$\beta_{ec\parallel}$ solar wind being more unstable with respect to the kinetic instabilities. We show that the core and the halo components for the high-$\beta_{ec\parallel}$ solar wind sometimes appear close to the whistler anisotropy instability, giving way to the generation of sunward propagating whistlers, which can resonate and scatter the strahl electrons.

For now the available in-situ observations only reach down to 0.3 au, but to globally understand the interplay between collisions close to the Sun, and then focusing and scattering of the strahl electrons along their expansion, we need to probe the regions even below the mentioned distance from the Sun. Therefore, a combination of numerical solar wind simulations and the soon available Parker Solar Probe data might be the key to a better understanding of the kinetic properties of the solar wind electrons. In the scope of this article we used the available observations to estimate the strahl PAW at 0.16 au, a distance of the first Parker Solar Probe perihelion. Obtained results point to the fading of the correlation between the strahl PAW and electron energy, with the PAWs in the low-$\beta_{ec\parallel}$ solar wind of $\sim$ 29$^o$, and in the high-$\beta_{ec\parallel}$ solar wind ranging between 37$^o$ and 65$^o$.

\section*{Acknowledgements}

We acknowledge all members of the Helios data archive team (\url{http://helios-data.ssl.berkeley.edu/team-members/}) to make the Helios data publicly available to the space physics community. We thank to D. Stansby for making available the new Helios proton core data set (\url{https://doi.org/10.5281/zenodo.891405}).
This work was supported by the Programme National PNST of CNRS/INSU co-funded by CNES.

All the analysis was done, and the plots produced using open source Python libraries NumPy, Matplotlib, Pandas, and SciPy.

We are grateful for the reviewers comments which were constructive and helped to improve the quality of the present work.

%%%%%%%%%%%%%%%%%%%%%%%%%%%%%%%%%%%%%%%%%%%%%%%%%%

%%%%%%%%%%%%%%%%%%%% REFERENCES %%%%%%%%%%%%%%%%%%

% The best way to enter references is to use BibTeX:

%\bibliographystyle{mnras}
%\bibliography{example} % if your bibtex file is called example.bib

% Alternatively you could enter them by hand, like this:
% This method is tedious and prone to error if you have lots of references
\bibliography{scatt.bib}

%%%%%%%%%%%%%%%%%%%%%%%%%%%%%%%%%%%%%%%%%%%%%%%%%%

%%%%%%%%%%%%%%%%% APPENDICES %%%%%%%%%%%%%%%%%%%%%

%\appendix

%\section{Some extra material}

%If you want to present additional material which would interrupt the flow of the main paper,
%it can be placed in an Appendix which appears after the list of references.

%%%%%%%%%%%%%%%%%%%%%%%%%%%%%%%%%%%%%%%%%%%%%%%%%%

% Don't change these lines
%\bsp	% typesetting comment
%\label{lastpage}
\end{document}